\documentclass{article}

\usepackage{microtype}
\usepackage{graphicx}
\usepackage{subfigure}
\usepackage{booktabs} 

\usepackage{hyperref}

\newcommand{\eg}{\hbox{\emph{e.g.,}}\xspace}
\newcommand{\ie}{\hbox{\emph{i.e.,}}\xspace}
\newcommand{\wrt}{\hbox{\emph{w.r.t.}}\xspace}

\usepackage[accepted]{icml2024}

\usepackage{amsmath}
\usepackage{amssymb}
\usepackage{mathtools}
\usepackage{amsthm}
\usepackage{xspace}
\usepackage{multicol}
\usepackage{multirow}
\usepackage{longtable}
\usepackage{inconsolata}
\usepackage{enumitem}
\usepackage{xfrac}
\usepackage{tikz}
\usetikzlibrary{calc}
\newcommand{\tikzmark}[1]{\tikz[overlay,remember picture] \node (#1) {};}
\usepackage{listings}
\usepackage[capitalize,noabbrev]{cleveref}

\theoremstyle{plain}

\theoremstyle{definition}

\theoremstyle{remark}

\usepackage{soul}
\usepackage{color}
\newcommand{\hlc}[2][yellow]{{%
    \colorlet{foo}{#1}%
    \sethlcolor{foo}\hl{#2}}%
}
\usepackage{makecell}
\usepackage{pifont}
\newcommand{\cmark}{\ding{51}}%
\newcommand{\xmark}{\ding{55}}%
\newcommand{\rtcexmatch}{\ensuremath{\RTC_{\text{ExactMatch}}}\xspace}

\DeclareMathOperator{\similarity}{sim}

\newcommand{\RTC}{\textsf{RTC}\xspace}

\icmltitlerunning{Unsupervised Evaluation of Code LLMs with Round-Trip Correctness}

\begin{document}

\twocolumn[
\icmltitle{Unsupervised Evaluation of Code LLMs with Round-Trip Correctness}

\icmlsetsymbol{equal}{*}

\begin{icmlauthorlist}
\icmlauthor{Miltiadis Allamanis}{equal,gdm}
\icmlauthor{Sheena Panthaplackel}{equal,gdm}
\icmlauthor{Pengcheng Yin}{equal,gdm}
\end{icmlauthorlist}

\icmlaffiliation{gdm}{Google DeepMind}

\icmlcorrespondingauthor{Miltiadis Allamanis}{mallamanis@google.com}

\icmlkeywords{Code LLM, evaluation}

\vskip 0.3in
]



\printAffiliationsAndNotice{\icmlEqualContribution} 

\begin{abstract}
To evaluate large language models of code, research has relied on a few small manually curated benchmarks, such as HumanEval and MBPP, which represent a narrow part of the real-world software domains. In this work, we introduce round-trip correctness (\RTC) as an alternative evaluation method.
\RTC allows Code LLM evaluation on a broader spectrum of real-world software domains without the need for costly human curation.
\RTC rests on the idea that we can ask a model to make a prediction (\eg describe some code using natural language), feed that prediction back (\eg synthesize code from the predicted description), and check if this round-trip leads to code that is semantically equivalent to the original input.
We show how to employ \RTC to evaluate code synthesis and editing.
We find that \RTC strongly correlates with model performance on existing narrow-domain code synthesis benchmarks while allowing us to expand to a much broader set of domains and tasks which was not previously possible without costly human annotations.

\end{abstract}

\section{Introduction}

While large language models (LLMs) have shown exceptional abilities in a wide range of tasks, their evaluation remains costly, commonly requiring laborious human-curated datasets.
This is particularly true for code capabilities of LLMs that commonly require highly-skilled programmers to create evaluation benchmarks.
Existing benchmarks, such as HumanEval~\citep{chen2021evaluating}, MBPP~\citep{austin2021program}, ARCADE~\citep{yin2022natural}, and DS-1000~\citep{lai2023ds} have been developed by asking human annotators to provide natural language, code, tests, and sometimes the target problems themselves.

At the same time, established human-annotated evaluation benchmarks focus on narrow domains of code:
HumanEval and MBPP evaluate introductory-level, standalone programming tasks; ARCADE and DS-1000 focus on simple data science tasks using few popular open-source libraries (\eg \texttt{numpy}, \texttt{pandas}).
These benchmarks are ``necessary'' but not ``sufficient'' for good performance for Code LLM users that work on different development environments which feature a much broader spectrum of domains, programming libraries and frameworks.
However, manually creating new benchmarks with expanded scope is costly and impractical given the general-purpose nature of code and fast-paced software evolution.

Towards ameliorating these limitations, we use the concept of round-trip correctness (\RTC) which allows us to perform unsupervised evaluation over certain LLM capabilities, complementing existing human-annotated evaluations.
\RTC relies on the idea that we can use an LLM to perform both an action (\eg \emph{``Describe these lines of code''}) and its inverse (\emph{``Implement the code given this description''}). See \autoref{fig:rtc synthesis} for an example.
We can then judge whether the round-trip has retained the semantics of the input.
This can be achieved via computing various discrete (\eg exact match) or continuous metrics (\eg BLEU) or involve execution-based oracles (\eg unit tests) which commonly already exist and require no additional human involvement.
Our contributions are:
\begin{itemize}[nosep, leftmargin=1em]
    \item We propose an unsupervised method for evaluating LLMs via \emph{round-trip correctness} (\RTC) (\autoref{sec:rtc def}) and instantiate it for code synthesis and editing (\autoref{sec:csc for code}).
    \item We show that \RTC strongly correlates with existing metrics on narrow-domain benchmarks (HumanEval and ARCADE) measuring the same LLM capability within that narrow domain (\autoref{subsec:rtc correlation}).
    \item We show that \RTC allows us to measure an LLM's performance over a wide-range of real-life software domains --- without human-provided annotations --- and complements existing narrow-domain benchmarks (\autoref{subsec:rtc domain variance}).
    \item We demonstrate that \RTC is a general metric that can be used for other tasks, like code editing, for which there are no well-established benchmarks (\autoref{subsec:editing rtc}).
\end{itemize}
Our code can be found at \url{https://github.com/google-deepmind/icml2024-roundtrip-correctness}.
\section{Round-Trip Correctness}
\label{sec:rtc def}

\paragraph{Background}
We draw inspiration from a software testing technique known as property-based testing~\citep{fink1997property}.
It allows defining properties that must hold between inputs and outputs of a program (\eg all items in the input list must also appear in the output list).
Round-trip correctness is one such property (\eg compressing and subsequently decompressing data must yield the original data).
Property-based testing allows software developers to expand their tests beyond few human-curated examples.
In this work, we re-purpose this concept for LLM evaluation.

\paragraph{\RTC for Model Evaluation}
\begin{figure}
    \centering
    \includegraphics[trim={1.1cm 0.45cm 1.9cm 0.55cm},clip,width=\columnwidth]{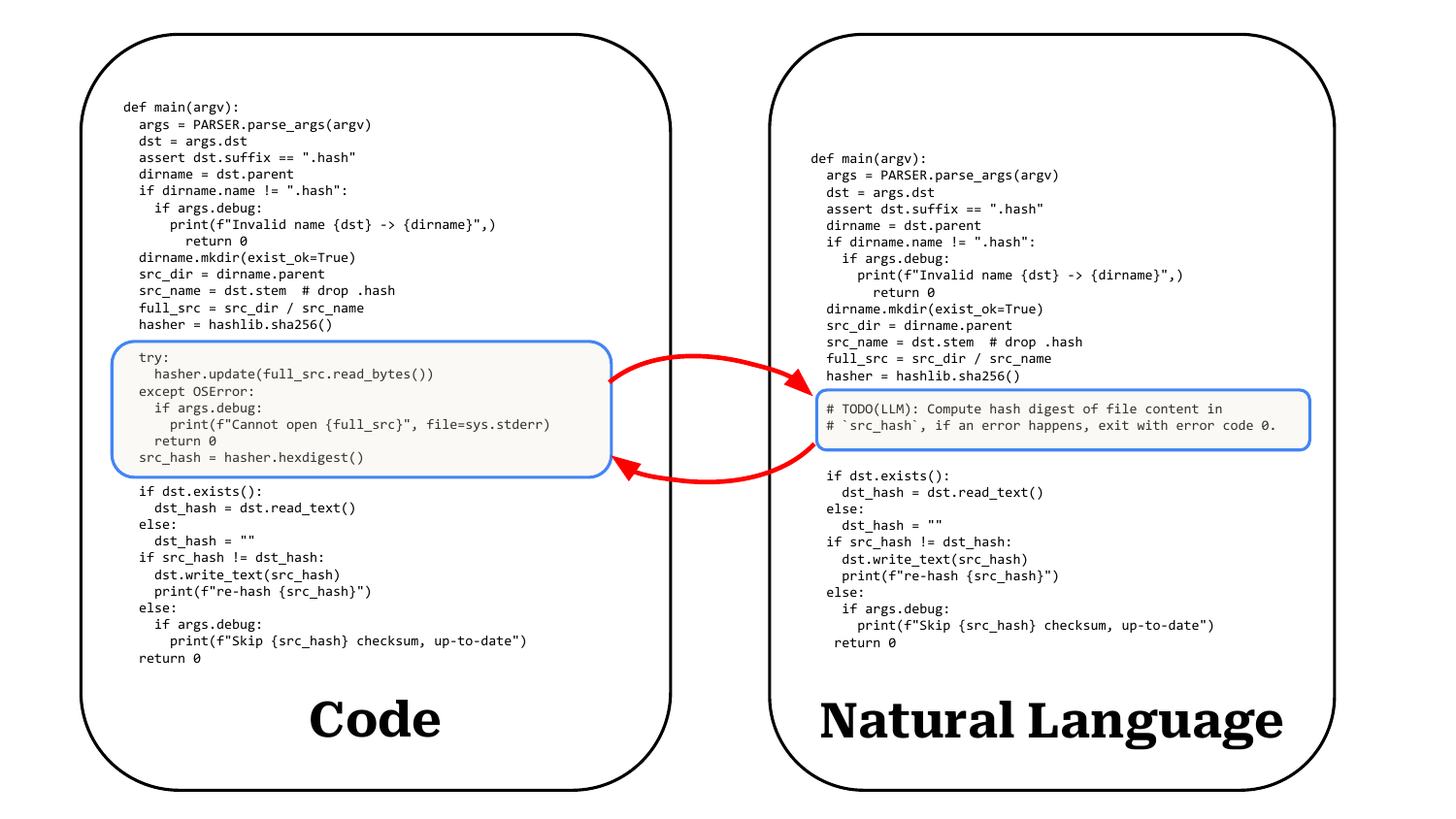}\vspace{-1em}
    \caption{Round-trip correctness (\RTC) for Code Synthesis: An LLM is asked to describe the highlighted code (left) within the context of the file. Subsequently, it is asked to implement the relevant code within the code context given the description it previously generated (right).}\label{fig:rtc synthesis} \vspace{-1em}
\end{figure}
\newcommand{\fwModel}{\ensuremath{M}\xspace}
\newcommand{\bwModel}{\ensuremath{M^{-1}}\xspace}
Consider two forms of data $\mathbb{X}$ and $\mathbb{Y}$, such as source code and natural language descriptions of code (\autoref{fig:rtc synthesis}) and
two (probabilistic) models whose task is to ``translate'' from one form of data to the other, \ie a forward model $\fwModel:\mathbb{X}\rightarrow\mathbb{Y}$ and a backward model $\bwModel:\mathbb{Y}\rightarrow\mathbb{X}$.
These models could be a single LLM prompted differently.

The central idea for unsupervised evaluation is the concept of \emph{round-trip correctness} (\RTC).
Intuitively, for a ``good'' forward and backward model we expect $\hat{x}=\bwModel(\fwModel(x))$ to be semantically equivalent to $x$.
For example, as we discuss in \autoref{sec:csc for code}, we can describe code with natural language in the forward pass and then generate back the code from the sampled natural language descriptions in the backward pass (\autoref{fig:rtc synthesis}).
To compute \RTC we need some function $\similarity(x, \hat{x})$ that estimates the semantic equivalence between the original $x$ and each predicted sample $\hat{x}$.
Such functions may include discrete or continuous metrics such as exact match, CodeBLEU~\citep{ren2020codebleu}, or CodeBERTScore~\citep{zhou2023codebertscore}, and execution-based semantic equivalence oracles, such as unit test execution.

We can then measure round-trip correctness as the ability of \fwModel and \bwModel to accurately perform the round-trip from an $x\in X$ to a $y = \fwModel(x) \in \mathbb{Y}$ and back to $\hat{x} = \bwModel(y) \in X$,
\begin{equation}\label{eq:csc abstract}
    \RTC_{\similarity}(x) \triangleq E_{y\sim \fwModel(x)}\left[E_{\hat{x}\sim \bwModel(y)}\left[\similarity(\hat{x}, x)\right]\right],
\end{equation}
where $\similarity(\cdot)$ estimates the semantic equivalence of $x$ and $\hat{x}$.
Since we cannot compute the expectations exactly, we draw small number of $N_f$ forward and $N_b$ backward samples and approximate \RTC as
\begin{equation*}
    \RTC_{\similarity}(x) \approx \frac{1}{N_f N_b}\sum_{y \sim \fwModel(x)} \sum_{\hat{x}\sim \bwModel(y)}\similarity(\hat{x}, x).
\end{equation*}

\paragraph{Measuring the forward lift}
In certain situations, it is possible for the backward model \bwModel to perform its task without any input from the forward model or with very low-quality forward model samples.
For instance, in text-to-code synthesis (\autoref{fig:rtc synthesis}), the code may be obvious within the code context, without requiring any explicit natural language instruction.
To measure this, we employ the notion of the \textit{forward lift}. Namely, we replace the forward sample with an uninformative/generic utterance $\epsilon$ and measure it as 
\newcommand{\lift}{\ensuremath{L_\fwModel}\xspace}
\newcommand{\liftsim}{\ensuremath{\lift^{\similarity}}\xspace}
\begin{equation*}
    \liftsim(x) = \RTC_{\similarity}(x) - E_{\hat{x}\sim \bwModel(\epsilon)}[\similarity(\hat{x}, x)].
\end{equation*}
$\liftsim$ can also be approximated by sampling.
Thus, \liftsim acts as a measure of the additional information encoded in the forward samples $\{ y \}$ that is available to the backward model \bwModel and can serve as a weak measure of a model's ability to perform the forward task.
A negative \liftsim may imply that the forward model \fwModel makes ``confusing'' predictions, distracting the backward model from performing its task.
A lift larger than zero implies that \fwModel provides helpful predictions that contribute towards the backward generation.
Finally a \liftsim close to zero may imply that the forward model \fwModel yields uninformative predictions or the input samples $x$ are easy for the backward model and \fwModel cannot possibly provide any additional information.

\paragraph{Limitations}
While \RTC allows us to evaluate Code LLMs without human annotations, it is not without limitations.
First, the quality of \RTC as a measure depends on that of the similarity function $\similarity(\cdot)$.
A weak measure of semantic similarity may yield arbitrary results.
Second, the measurement of the performance of the forward and backward tasks is coupled: if \fwModel is unable to provide plausible samples, we cannot expect to measure the ability of the backward model \bwModel. This may be a problem if we care for \emph{only one} of the forward or backward tasks.

Finally, \RTC assumes ``reasonably'' trained and instruction-tuned LLMs. In an adversarial setting, a forward model \fwModel can ignore the instruction and recite its input $x$.
Then, the backward model \bwModel can copy the output of the forward model \fwModel achieving perfect \RTC.
While this is unlikely to happen for models that employ common (pre)training methods, if we train/fine-tune models with the objective of \autoref{eq:csc abstract} such a behavior can arise naturally.

\section{\RTC for Code}
\label{sec:csc for code}

While \RTC is general, it is well-suited for code where automatically judging the semantic similarity (via $\similarity(\cdot)$) is easier than natural language, commonly through proxies such as unit tests.
In this section, we discuss two code-specific evaluations for two common code application: code synthesis and code editing.
A summary is shown in \autoref{tbl:code task summary}.

\newcommand{\synthesisrtc}{\textsc{SynthesisRtc}\xspace}
\newcommand{\editingrtc}{\textsc{EditingRtc}\xspace}

\begin{table*}
\caption{Summary of Code \RTC tasks discussed in this work.}\label{tbl:code task summary}
\begin{tabular}{lll} \toprule
     & \synthesisrtc & \editingrtc \\ \midrule
    Input $\mathbb{X}$ & Source Code Region (one or more statements) & Code Edit  \\
    Intermediate $\mathbb{Y}$ & Natural Language Description of Code Region & Natural Language Description of Edit \\
    Forward Task Context & Source Code around Target Region & Source Code around Edit\\
    Backwards Task Context & Source Code around Target Region & Original Version Code before Edit \\
    Uninformative utterance $\epsilon$  & Code Completion/Infilling of Target Region & -- \\
    $\similarity(\cdot)$ & All Unit Test Pass (0/1) & Exact Match\\
      \bottomrule
\end{tabular}\vspace{-1em}
\end{table*}

\subsection{Round-trip Code Synthesis (\synthesisrtc)}
\label{sub:synthesisrtc}
Code synthesis with LLMs is one of the most studied tasks.
Evaluating on this task commonly requires a human-annotated dataset of natural language descriptions, function signatures (or full implementations), and unit tests.
Instead, for \RTC we construct \synthesisrtc as an in-context code synthesis task (\autoref{fig:rtc synthesis}) that does \emph{not} require input natural language descriptions.
Given a coherent region of code within a source code file, we ask a forward LLM (\fwModel) to describe the code region concisely with natural language.
This yields a natural language utterance $y$.
Then, we remove the code region and replace it with a \texttt{TODO} comment containing the natural language description generated by \fwModel.
Finally, we ask \bwModel to synthesize code implementing the \texttt{TODO}.

For a model to be round-trip correct, the synthesized code must be semantically equivalent to the original.
\newcommand{\rtcpass}{\ensuremath{\RTC_{\text{pass}}}\xspace}
In practice, proving semantic equivalence is hard, and we employ unit tests as a proxy and measure \rtcpass.
Note that unit tests are commonly included in well-developed code, and so they are often readily available without any additional human effort.
If they do not exist, automatic test generation methods~\citep{nebut2006automatic} can be employed, but it is also possible to consider other weaker similarity functions like exact match or BLEU, as done in prior work.
Finally, to measure the lift for \synthesisrtc we can use an uninformative \texttt{TODO} comment (\eg \emph{$\epsilon=$``\texttt{TODO: Implement.}''}) and check the correctness with $\similarity(\cdot)$.
The lift $\lift^{\text{pass}}$ can be attributed to the additional information provided by the forward model \fwModel.

\subsection{Round-trip Code Editing (\editingrtc )}
\label{subsec:editing rtc}
In addition to \synthesisrtc, which evaluates an LLM's ability to generate new code from scratch, we also explore \RTC as an evaluation metric for code editing.
Editing is a common real-world software engineering scenario in which code is modified (\eg bug fixing, refactoring, implementing new or updated functionality). Performing this task requires complex reasoning, which entails identifying parts of the existing code that need to be changed, determining how they should be changed, and retaining all other parts of the code~\citep{jimenez2023swe}. However, this is not a common evaluation task for LLMs, and there are no well-established metrics or benchmarks.

Our approach closely follows \autoref{sub:synthesisrtc}.
Namely, as the input to the forward model \fwModel we provide an edit, represented as an \textit{old} code snippet followed by the \textit{new} version (with the edit applied). We ask the model to describe the edit concisely using natural language, which results in a predicted \textit{edit description} $y$. Finally, we provide the old code snippet (before the edit) and the generated edit description to the backward model \bwModel, and we ask it to generate the new code, \ie, apply the edit described in $y$.

Rather than relying on human-provided natural language labels to perform supervised evaluation, we leverage \RTC, in which we compute the similarity between the original edit that is provided as input to \fwModel and the edit predicted by \bwModel. Similar to \citet{li2022automating}, we use exact match as the similarity metric $\similarity(\cdot)$.\footnote{If unit tests were available we would have preferred them.} To approximately measure the quality of the generated edit descriptions we compute the lift for a baseline task in which we provide \bwModel an uninformative edit description ($\epsilon=$``\texttt{Edit.}'').

\section{Evaluation}

\paragraph{Experimental Setup}
Unless stated otherwise, to compute \RTC we draw 3 forward samples and one backward sample per forward sample.
We use temperature of 0.8 for the forward model (to allow for disparate forward samples) and 0.1 for the backward samples (to generate high-probability code generations).
We use three-shot prompting with identical few-shot prompts for all models.
Given the time constraints and compute limitations, we selected these hyperparameters and have not explored any variations.
Finally, we limit the length of the forward samples to 128 characters.
In this way, models are forced to ``compress'' their understanding in a succinct natural language sentence and more verbose models cannot gain an advantage over less verbose ones.

We note that  different forward and backward models could be used.
However, there are confounders in doing so that we wanted to avoid: different forward models may generate natural language descriptions that are confusing to different backward models. Such a communication chasm would misrepresent the capabilities of these models.
For instance, consider two models $M_1$ and $M_2$ that are both good at code synthesis based on natural language instructions.
However, suppose $M_1$ performs poorly in code-to-natural language generation whereas $M_2$ is very good. Now, if we use $M_1$ to create forward samples for the \synthesisrtc task, we may end up with low-quality natural language descriptions of code. If we then provide these low-quality descriptions as input to $M_2$ to generate backward samples, we may also end up with low-quality code samples. This would lead us to falsely conclude that $M_2$ performs poorly on code synthesis.
Due to these, we aim to minimize any extrinsic factors that may affect the result and use the same model for both the forward and backward samples to compute \RTC.
At the same time, using the same model for both the forward and backward tasks can also enable \RTC to measure a model’s skill at generating both code as well as natural language targets. This would help distinguish more ``well-rounded'' LLMs that excel at both code and description generation from models that can only handle a single task (\eg models trained on code-heavy data may only predict code, see \citet{muennighoff2023octopack}).

\subsection{Does \RTC correlate with existing metrics on narrow-domain benchmarks?}
\label{subsec:rtc correlation}
We make the hypothesis that \RTC, when applied to existing benchmarks, is highly correlated with widely accepted metrics, such as pass@$k$~\citep{chen2021evaluating} which in turn is recognized as a metric strongly correlating with how LLM users perceive its performance.

To test this hypothesis we employ HumanEval~\citep{chen2021evaluating} and ARCADE~\citep{yin2022natural} that represent two common domains: general-purpose coding to solve algorithmic problems and multi-turn data science programming in Jupyter notebooks.
HumanEval is a function-level code generation dataset where each problem is a Python function with a natural language description in its docstring.
We re-purpose each problem by removing the docstring and use the ground-truth function solution as input to the forward model.
LLMs are subsequently asked to describe the body of the ground-truth function (forward) and then implement it based on the generated description.
For ARCADE, each original problem comes with a natural language question (\eg \textit{``What's the GDP growth rate of that country?''}) and a ground-truth code solution along with any notebook context (\eg the question and solution for a prior turn \textit{``Which country received the highest aid?''}). To compute \RTC, we use the notebook context and the code solution of a turn as input to generate question descriptions $y$ in the forward pass.
The backward model is subsequently asked to implement the code using the generated description $y$ and the notebook context. The few-shot prompts of \citet{yin2022natural} are used. 

\begin{table}
    \centering
    \caption{\rtcpass \emph{vs} standard pass@1 metric and $\lift^{\text{pass}}$ across models. Results sorted by \rtcpass. DSC stands for DeepSeekCoder.}\label{tbl:correlation}
    \begin{tabular}{llrrr} \toprule
         && pass@1 & \rtcpass  & $\lift^{\text{pass}}$\\ \midrule
    \multicolumn{4}{l}{\textbf{HumanEval}~\citep{chen2021evaluating}} \\
    &PaLM 2-S      & 19.5\%  & 8.3\%  & -0.2\%\\
    &PaLM 2-S+     & 29.3\%  & 10.6\% & 3.3\%\\
    &PaLM 2-S*     & 37.6\%  & 18.3\% & 9.8\%\\ 
    &Gemini Nano 2 & 33.4\%  & 18.9\% & 3.7\%\\
    &StarCoder2 15B & 46.3\% & 31.7\% & 20.7\% \\
    &Gemini v1 Pro & 63.4\%  & 34.8\% & 19.6\%\\ 
    &DSC33B-IT & 75.6\% & 40.2\% & 30.4\% \\\midrule
    \multicolumn{4}{l}{\textbf{ARCADE}~\citep{yin2022natural}} \\
    &PaLM 2-S      & 5.5\% & 2.7\%  & --\\
    &PaLM 2-S+     & 8.2\% & 3.5\%  & --\\
    &PaLM 2-S*     & 15.3\% & 6.5\%  & --\\ 
    &Gemini Nano 2 & 14.4\% & 7.7\% & --\\
    &Gemini v1 Pro    & 18.3\% & 11.1\% & -- \\ 
    &StarCoder2 15B & 25.4\% & 12.1\% & -- \\
    &DSC33B-IT &  24.8\% & 15.1\%  & -- \\
    \bottomrule
    \end{tabular}\vspace{-1em}
    
\end{table}

\autoref{tbl:correlation} show the scores achieved by two classes of models PaLM~2~\citep{anil2023palm} and Gemini~\citep{team2023gemini}
along with two open models DeepSeekCoder-33B-Instruct~\citep{guo2024deepseek} (abbreviated as as DSC33B-IT) and StarCoder2~\citep{lozhkov2024starcoder}.
The HumanEval pass@1 scores for Gemini, StarCoder2, and DeepSeekCoder models were obtained from the base scores of the independent \href{https://evalplus.github.io/leaderboard.html}{EvalPlus Leaderboard}.
While not surprising, we see that there is a strong correlation between the ``standard'' pass@1 and \rtcpass for both benchmarks.
The Pearson correlation is $r=0.96$ for HumanEval and $r=0.96$ for ARCADE and the Spearman correlation is $\rho=0.90$ and $\rho=0.81$ respectively.

These results suggest that \RTC strongly correlates with existing metrics on benchmarks that have been curated through costly human annotations.
Thanks to the strong correlation, we can conclude that \RTC is a valid metric that reflects the real-world performance of LLMs and thus can complement existing human-annotated benchmarks.
It should be noted that these correlations are not perfect.
This is to be expected as \rtcpass couples the code-to-text and the text-to-code synthesis capabilities whereas pass@1 only measures the text-to-code synthesis capability of each model.

One potential caveat could be that \RTC's correlation with pass@1 is only due to the different sizes of the models, as it is generally true that larger models overperform smaller ones.
However, all the PaLM 2 models used have the same number of parameters and \RTC and pass@1 still correlates strongly across these models. 
This suggests that \RTC correlates with the coding abilities controlling for model size rather than some other capability that coincidentally improves with model size.

Note that the absolute \rtcpass is worse than the pass@1 for HumanEval in absolute terms. 
This difference can be explained by two factors: (a) the standard HumanEval prompt includes input-output examples which the forward model is not prompted to generate.
This makes the backward code generation task harder.
Indeed, recently \citet{liu2024codemind} confirmed that removing those input-output examples from the original HumanEval prompts leads to significant drop in pass@1 \citep[Table 3]{liu2024codemind}.
(b) The forward description generation task also introduces some noise.

\paragraph{Evaluating Code-to-Description}\autoref{tbl:correlation} show the lift $\lift^{\text{pass}}$ for HumanEval.
For \synthesisrtc the uninfomative backward task is  code completion within the code context but no (natural language) information about the implementation or its intention.
For HumanEval, \autoref{tbl:correlation} shows that all models --- except from PaLM 2-S --- show a lift, but better models offer a larger lift than small ones.
For ARCADE, the baseline task (code completion without any natural language instruction) is \emph{not} meaningful given the nature of the benchmark as the baseline backward pass@1 would be always zero and hence we do not compute it. 

\paragraph{Sensitivity of \RTC}
Similar to pass@$k$ metrics LLM sampling temperature, number of samples per example, and dataset size affect the statistical robustness of the RTC reported. In this work, we chose to use low temperature for the backward model and a mildly high temperature for the forward model. The low temperature of the backward pass renders the $RTC_{pass}$ relatively stable: we performed the HumanEval experiments (\autoref{tbl:correlation}) 10 times and measure the standard deviation to be $\sigma=1.11\%$ which is small and shows that the results shown are fairly stable.
Changing $N_f$, $N_b$ and the forward/backward sampling temperatures beyond the ones discussed changes the results. For example, we observe that increasing the temperature of the backward model requires a significant increase in $N_b$ to reduce variance.

\subsection{Do LLMs perform similarly across domains?}
\label{subsec:rtc domain variance}
\begin{figure*}
\centering
\includegraphics[trim={3.5cm 0.5cm 3.7cm 0.5cm},clip,width=\textwidth]{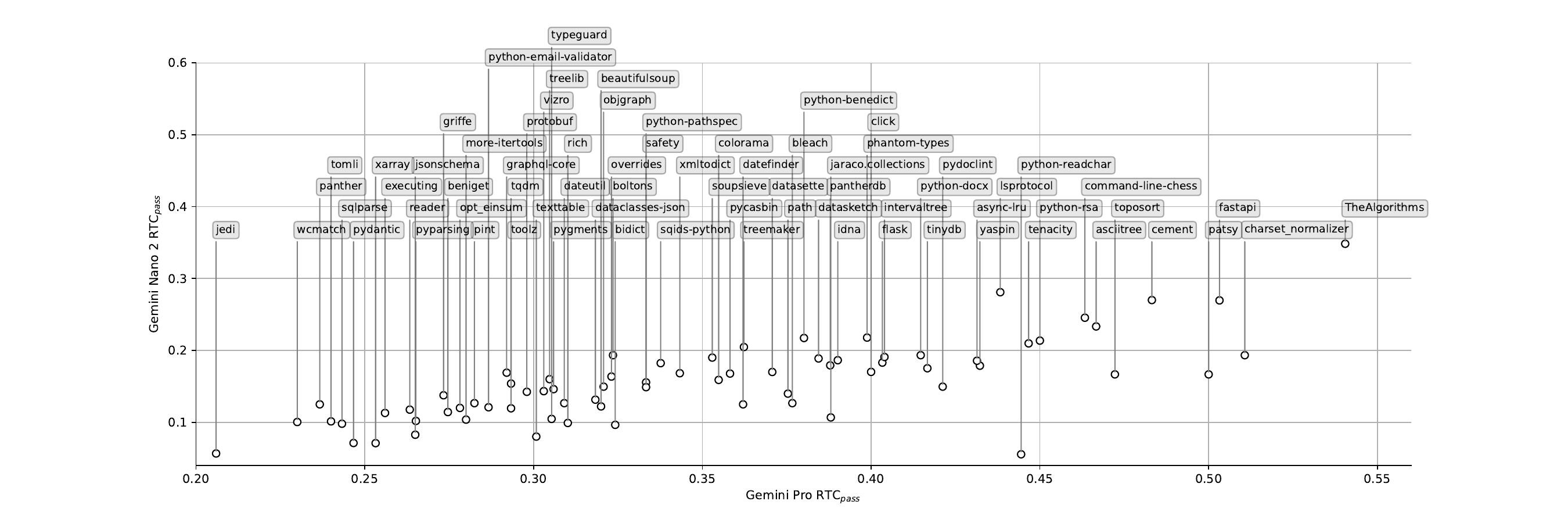}\vspace{-1em}
    \caption{Round-trip correctness (\RTC) for Code Synthesis across 58 open-source projects of diverse domains for Gemini Pro and Nano 2: \rtcpass varies widely across projects/domains, something that common code synthesis benchmarks fail to capture.}
    \label{fig:rtc circus}\vspace{-1em}
\end{figure*}

Previously, we showed that \RTC strongly correlates with pass@1 on limited-domain, human-annotated benchmarks.
It is unclear whether these narrow-domain benchmarks fully reflect an LLM's capabilities across all code domains.
To investigate this question, we collect a set of 77 of permissively licensed open-source Python projects that have a test suite which we can execute and all the tests pass.
We then sample ranges within the code by sampling one or more consecutive sequential statements from the project's code whose size is between 32 and 384 characters long and are covered by the test suite.
Specifically, each statement corresponds to a node on the concrete syntax tree (CST) of the code.
The probability of a sequence of statements being sampled is proportional to the number of characters they contain and inversely proportional to the number of other candidate nodes in the CST that also contain those characters.
Files containing unit tests are excluded.
We also filter out any ranges that when deleted have no effect to the results of the test suite.
This ensures that the selected code has at least some observable effect to the expected outputs.
This last step is crucial and filters about 40\% of the sampled ranges, although this statistic varies across projects.

Finally, we randomly sample 100 ranges per-project. If for a project we cannot collect 80 or more samples, we discard the entire project.
The resulting dataset is made of 5,961 samples from 58 open source Python projects representing varying domains.
To avoid giving advantage to models with longer contexts, we provide as context the few lines before and after each selected code range such that no more than 1024 characters of context are included and each line is either shown entirely or not at all.
Alternative implementations, \eg including the entire file, more files, or even an entire repository as in \citet{ding2024crosscodeeval} are also reasonable and we expect they will become increasingly viable in the future.
The used projects are shown in \autoref{appx:oss project list}. 

Subsequently, we compute \rtcpass as in \autoref{subsec:rtc correlation}.
\autoref{fig:rtc circus} shows a plot of the average \rtcpass for each project showing a wide variability on  both Gemini Nano 2 and Gemini Pro's performance.
This is not unexpected: HumanEval and ARCADE only test for two specific, fairly narrow domains that do not reflect the full diversity of domains in real-life software.
From \autoref{fig:rtc circus} we see that for some projects the performance is really good compared to the rest.
In particular, the \href{https://github.com/TheAlgorithms/Python}{\texttt{TheAlgorithms}} project that contains educational implementations of popular computer science algorithms achieves very good \rtcpass.
In contrast, \href{https://github.com/davidhalter/jedi}{\texttt{jedi}}, a static analysis library for Python has the lowest \rtcpass.
There is also a wide variability among the \rtcpass of the other projects showing the diverse performance of LLMs across different domains.
While the \rtcpass of the two models are correlated ($r=0.75$, Spearman's $\rho=0.76$), this correlation is not perfect, showing that different LLMs can have varying performance characteristics in different domains.

These results suggest that existing narrow-domain benchmarks do \emph{not} capture the LLM's capabilities across multiple domains.
Therefore for Code LLMs that need to support a wide range of coding domains additional benchmarks are needed.
While manually curating and annotating such benchmarks is possible, unsupervised evaluation through round-trip correctness offers a reasonable alternative.

\paragraph{Lift across domains}
\label{subsec:lift domain variance}
The lift $\lift^{\text{pass}}$ shows a similar variation across different projects (shown in \autoref{appx:lift plot} in \autoref{fig:lift circus}).
We note that the average lift for Gemini Nano 2 is $\lift=7.0\%$ and for the Gemini Pro $\lift=21.5\%$ which is higher compared to HumanEval (\autoref{tbl:correlation}).
This may be because HumanEval's function signatures used as context (\eg {\small \texttt{def is\_palindrome(string: str) -> bool}}) in the backward model are somewhat informative and the forward samples may \emph{not} be able to provide additional information. This further suggests that HumanEval may be a relatively simple benchmark compared to the one in this section.

\paragraph{Qualitative Analysis}
\begin{figure}[t!]
\centering
\begin{tabular}{p{\columnwidth}} \toprule
\textbf{Input Code and Target Region} \\ \midrule
\vspace{-1em}
\begin{lstlisting}[language=Python]
def unique_in_window(iterable, n, key=None):
  """Yield the items from *iterable* that haven't been
  seen recently. *n* is the size of the lookback window."""
  if n <= 0:
    raise ValueError('n must be greater than 0')

  window = deque(maxlen=n)
  counts = defaultdict(int)
  use_key = key is not None

(*@\tikzmark{r1}@*)  for item in iterable:
    if len(window) == n:
      to_discard = window[0]
      if counts[to_discard] == 1:
        del counts[to_discard]
      else:
        counts[to_discard] -= 1     (*@\tikzmark{r2}@*)

    k = key(item) if use_key else item
    if k not in counts:
      yield item
    counts[k] += 1
    window.append(k)
\end{lstlisting}
\tikz[overlay,remember picture]{
\draw [fill=yellow, fill opacity=0.1] (r1.south east) rectangle (r2.south west);
} \vspace{-2em}\\ \midrule
\textbf{Forward samples} \\ \midrule\footnotesize
\begin{enumerate}[noitemsep,topsep=0pt,parsep=0pt,partopsep=0pt,leftmargin=2pt]
    \item `to\_discard` takes the first value in `window`. If the count for that value in `counts` is `1`, we remove the value, otherwise we decrement the count.
    \item if the window is at capacity, discard the oldest element, and update counts so that this element is only considered if it is seen again.
    \item `to\_discard` equals `window[(len(window) - 1) \% len(window)]`
\end{enumerate} \vspace{-2em} \\ \midrule
\textbf{Backward samples} \\ \midrule
\vspace{-1.5em}
\begin{lstlisting}[belowskip=0em]

(*@\textsf{\textbf{Generation for forward sample \#1}}@*)
to_discard = window.popleft()
if counts[to_discard] == 1:
    del counts[to_discard]
else:
    counts[to_discard] -= 1

(*@\textsf{\textbf{Generation for forward sample \#2}}@*)
if len(window) == n:
    k = window.popleft()
    counts[k] -= 1
    if counts[k] == 0:
        del counts[k]

(*@\textsf{\textbf{Generation for forward sample \#3}}@*)
to_discard = window[(len(window) - 1) % len(window)]
\end{lstlisting}\vspace{-2em} \\ \bottomrule
\end{tabular}

    \caption{\synthesisrtc example from \texttt{\href{https://github.com/more-itertools/more-itertools/blob/8e4b048d9962e655d7b7f9bfc5e8b0675d70697f/more_itertools/more.py\#L4278}{more-itertools}} from Gemini Pro. Code slightly reformatted/abbreviated for space.}\label{fig:syntehsis example}\vspace{-1em}
\end{figure}

\autoref{fig:syntehsis example} shows a cherry-picked example from an open-source project that we have selected as a good representative of common error modes.
The target code range (yellow box) performs a relatively simple operation that maintains the invariant on the \texttt{counts} dictionary.
The first two natural language samples explain this in a plausible way.
However, the first sample misses the crucial information about the outer \texttt{if} condition, which is well-captured by the second one.
Overall, qualitatively we often see some important part of the logic not being captured in the natural language description.
Additionally, the first sample takes a very formulaic and unnatural way of describing the code giving emphasis on the low-level operations rather than the underlying logic.
This seems like a common error mode for the forward model and happens more often on shorter snippet.
The third sample seems like a hallucination and is just wrong.

All backward samples are reasonable implementations of their respective forward sample.
However, note that the second sample has the ambiguous phrase \emph{``discard the oldest element''} which can reasonably imply the use of \texttt{popleft()} instead of just peeking at the leftmost element of \texttt{window}.
Despite this difference and thanks to the implementation semantics of \texttt{deque} the second sample is semantically equivalent to the original code when placed within the code context, hence achieving an $\rtcpass=\sfrac{1}{3}$.

\subsection{Evaluating beyond Code Synthesis: Code Editing}

\begin{table}
    \caption{Evaluating \editingrtc with a random sample of 1K examples from the CodeReviewer test set of \citet{li2022automating}.}\label{tbl:editing-results}
\small
    \centering
    \begin{tabular}{lrr} \toprule
         & Gemini Nano 2 & Gemini Pro\\ \midrule
    \textbf{Unsupervised} && \\
     \rtcexmatch (\%)  & 5.2 & 12.9 \\
    $\lift^{\text{ExactMatch}}$ (\%) & 4.8 & 12.4  \\
     \ensuremath{\RTC_{\text{BLEU}}} (\%)  & 72.3 & 73.7 \\
     \ensuremath{\RTC_{\text{ROUGE}}} (\%)  & 81.4 & 82.3 \\
    \midrule
    \textbf{Supervised} && \\
    Edit $\rightarrow$ NL (BLEU) & 0.5 & 1.0 \\
    NL $\rightarrow$ Edit (Exact Match) & 11.7  & 19.4 \\
    \bottomrule
    \end{tabular}\vspace{-1em}
\end{table}

To evaluate code editing, we use the CodeReviewer test set~\cite{li2022automating}, which is a parallel corpus of GitHub Pull Request (PR) review comments paired with their edited code, including both the \textit{old} and \textit{new} versions (\ie, before and after the edit for resolving the review comment). We randomly sample 1K examples from the original test set. The resulting set contains samples from multiple programming languages (Java 17\%, Go 23\%, Python 21\%, C++ 9\%, PHP 7\%, JavaScript 7\%, Ruby 5\%, C\# 5\%, C 4\%). We use 3-shot prompting (with 3 examples from the CodeReviewer validation set) and conduct experiments with Gemini models~\cite{team2023gemini}. We draw 3 forward (temperature=1.0) and 1 backward sample (temperature=0.0).

Results are shown in \autoref{tbl:editing-results}. Overall, we observe trends consistent to those of \synthesisrtc (\autoref{tbl:correlation}): \rtcexmatch is higher for Gemini Pro, suggesting that Gemini Pro has superior capabilities in terms of generating natural language edit descriptions (forward task) and editing code based on natural language descriptions of edits (backward task). Additionally, the forward lift ($\lift^{\text{ExactMatch}}$) is much higher for Gemini Pro, once again demonstrating the quality of the descriptions generated by \fwModel.

Note that \rtcexmatch and $\lift^{\text{ExactMatch}}$ are computed in a completely unsupervised manner, without relying on any ground truth labels for edit descriptions or edited code snippets. We consider this to be particularly useful for benchmarking tasks related to code editing, as it is extremely difficult to collect a high-quality labeled test set. In fact, the CodeReviewer dataset was extracted automatically, using rough heuristics to align natural language edit descriptions with corresponding code edits. Consequently, it is a noisy test set, with some inconsistencies between descriptions and edits. If we take the PR comment as the ground truth edit description, we can perform standard evaluation on edit description generation (\ie, given the old and new versions of the code snippet, sample an edit description with temperature=0.0) using standard text generation metrics like BLEU~\citep{papineni2002bleu}. As shown in \autoref{tbl:editing-results}, these scores are very low for the two models and uninformative. 
Since in editing tasks much of the input is often preserved in the target output, a simple baseline which retains the input unchanged fares well with standard metrics like BLEU, ROUGE, and even embedding-based metrics like BERTScore~\citep{zhang2019bertscore}.
For this reason, we consider exact match to be a more reliable metric, but in \autoref{tbl:editing-results} we have included the other metrics for completeness.

One of the main reasons for this is that traditional supervised metrics for evaluating text generation rely on lexical overlap, against reference texts. Here, there is only one reference (\ie the PR comment) and the lexical overlap can be fairly limited. For instance, the PR comment in \autoref{edit-rtc-gemini-m-12} appears to be irrelevant to the edit and is thus of low quality. On the other hand, the three edit descriptions sampled from \fwModel are all arguably consistent with the edit. In fact, when they are fed into \bwModel, they lead to a predicted edit which exactly matches the original, resulting to an $\rtcexmatch=100.0\%$. However, the average BLEU score \wrt the PR comment is extremely low, at 1.837, which incorrectly suggests that the model fails to generate a high-quality edit description for this example.

Furthermore, we can perform supervised evaluation on code edit generation. That is, given the old version of the code snippet and the PR comment, we sample one edit with temperature=0.0 (greedy decoding). We can then compute exact match performance with respect to the ground truth code edit snippet in the CodeReviewer test set, as shown in \autoref{tbl:editing-results}. However, when the PR comment is misaligned with the labeled code edit, this evaluation will mischaracterize the code editing capabilities of an LLM. For the example in \autoref{edit-rtc-gemini-m-12}, the edit predicted based on the PR comment actually appears to be consistent with the comment. However, the labeled ground truth edit is inconsistent with the PR comment, and so the fact that the predicted edit does not match the ground truth edit will falsely suggest that the model fails to generate a high-quality edit for this example. This demonstrates that even when weakly labeled test sets are available, noisy labels and other drawbacks of supervised metrics often make the evaluation unreliable. In such cases, we believe that \RTC provides a more reliable evaluation. We show more \editingrtc examples in \autoref{app:editingrtc_examples}.

\begin{figure*}[ht!]
\small
\begin{tabular}{lcc}
\toprule
\begin{lstlisting}
Describe concisely and accurately with natural language the differences 
between the old and new code shown below.

[old]
 def remote_shutdown(self):
    udp_sock = socket.socket(socket.AF_INET, socket.SOCK_DGRAM)
    udp_sock.sendto(b"Shutdown", ((*@\hlc[pink]{"localhost"}@*), self.management_port))
    self.log.debug("Shutdown command sent")
    time.sleep(10)
    udp_sock.sendto(b"StopTestNow", ((*@\hlc[pink]{"localhost"}@*), self.management_port))
    self.log.debug("StopTestNow command sent")

[new]
 def remote_shutdown(self):
    udp_sock = socket.socket(socket.AF_INET, socket.SOCK_DGRAM)
    udp_sock.sendto(b"Shutdown", ((*@\hlc[green]{"127.0.0.1"}@*), self.management_port))
    self.log.debug("Shutdown command sent")
    time.sleep(10)
    udp_sock.sendto(b"StopTestNow", ((*@\hlc[green]{"127.0.0.1"}@*), self.management_port))
    self.log.debug("StopTestNow command sent")

[edit description]

\end{lstlisting}& \\
\midrule
\bf Description & \bf Edit $\rightarrow$ NL (BLEU) & \bf NL $\rightarrow$ Edit (Exact Match) \\
\hline
\small{\textbf{PR Comment:} After first command, check if JMeter has finished} & - & \xmark \\
\begin{lstlisting}
[new]
 def remote_shutdown(self):
    udp_sock = socket.socket(socket.AF_INET, socket.SOCK_DGRAM)
    udp_sock.sendto(b"Shutdown", ("localhost", self.management_port))
    self.log.debug("Shutdown command sent")
    time.sleep(10)
    udp_sock.sendto(b"StopTestNow", ("localhost", self.management_port))
    self.log.debug("StopTestNow command sent")
    (*@\hlc[green]{if self.is\_jmeter\_finished():}@*)
       (*@\hlc[green]{self.log.debug("JMeter has finished")}@*)
    (*@\hlc[green]{else:}@*)
       (*@\hlc[green]{self.log.debug("JMeter has not finished")}@*)
\end{lstlisting} && \\
\hline
\small{\textbf{Baseline:} Edit.}  &  - &  \xmark \\
\begin{lstlisting}
[new]
 def remote_shutdown(self):
    udp_sock = socket.socket(socket.AF_INET, socket.SOCK_DGRAM)
    udp_sock.sendto(b"Shutdown", ((*@\hlc[pink]{"localhost"}@*), self.management_port))
    self.log.debug("Shutdown command sent")
    time.sleep(10)
    udp_sock.sendto(b"StopTestNow", ((*@\hlc[pink]{"localhost"}@*), self.management_port))
    self.log.debug("StopTestNow command sent")
\end{lstlisting}&& \\
\hline
\textbf{\editingrtc Forward Samples:} && \\
\makecell[l]{\small{$\bullet$ Replace localhost with 127.0.0.1 to avoid potential conflicts}\\\small{on a dual-stacked machine.}} & 1.314 & \cmark \\
\small{$\bullet$ Use the constant of 127.0.0.1 instead of "localhost"} & 2.098 & \cmark \\
\small{$\bullet$ Please replace "localhost" with "127.0.0.1".} & 2.098 & \cmark\\
\midrule
& \bf Avg BLEU & \bf \rtcexmatch   \\
\hline
& 1.837 & 100.0 \\
\bottomrule
\end{tabular}
\caption{A CodeReviewer example with Gemini Pro predictions: 3 sampled descriptions in the forward pass as well as their corresponding predicted edits in the backward pass. We additionally include predictions from the backward pass when the provided description is instead the PR comment or baseline description. Examples have minor edits/re-format due to space constraints.} \label{edit-rtc-gemini-m-12}\vspace{-1em}
\end{figure*}

\section{Related Work}
\textbf{Supervised Code Evaluation Metrics} Accuracy, or how often model-generated code exactly matches reference code, is a metric for evaluating code generation~\citep{agashe2019juice, li2022automating, yin2017syntactic, chakraborty2021multi}. However, it is overly conservative and significantly underestimates the actual performance, as it is possible to generate semantically correct code without exactly matching a reference (\eg different variable names, ordering of statements, or logic). \citet{agashe2019juice, li2022automating, yin2017syntactic,wei2019code} among others adopted BLEU~\cite{papineni2002bleu}.
However, \citet{ren2020codebleu} showed that BLEU is not well-suited for evaluating code correctness since it fails to capture the syntax, semantics, and functionality of code. CodeBLEU~\cite{ren2020codebleu} addresses this by augmenting BLEU with syntactic and data flow information. However, CodeBLEU requires the model-generated code to follow the same structure of the reference code, which is still overly conservative penalizing correct code that follows different ordering of statements or logic while still partly relying on BLEU's token overlap score. More recently, \citet{zhou2023codebertscore} proposed CodeBERTScore which measures the similarity of the model-generated code and the reference code based on the dot-product similarity of their contextualized vector representations from a pretrained LLM. However, CodeBERTScore does not explicitly evaluate semantic similarity. 

To capture code's functional correctness, research has moved towards test-based oracles: generated code is executed against predefined test cases, and if the generated code passes these test cases, it is considered correct. This does not require the model-generated code to match the naming, structure, or logic of a reference, but checks if aspects of the functionality are correct. It should be noted that unit tests are partial indicators of functional correctness and cannot guarantee semantic equivalence which in the general case is undecidable.

\textbf{Self-Consistency}
Self-consistency is often used to describe how consistent a model's generated samples are with one another. If a model generates consistent predictions multiple times for the same input, the model is likely more confident that those predictions are correct. Based on this intuition, \citet{wang2022self} designed a decoding strategy using self-consistency to identify the most likely correct answer from a set of samples, which they showed improved the model's ability to do chain-of-thought reasoning. While self-consistency can be used as an uncertainty estimator, it is not always well-suited to evaluate accuracy, as a model can be consistently wrong. In contrast, \RTC can be a more reliable metric since it requires functional correctness which is a stronger signal than consistency.

\RTC also relates to IdentityChain of \citet{min2023beyond} who propose to measure the self-consistency of a Code LLM via multiple efforts to make a round-trip. In contrast to \RTC, IdentityChain still requires an annotated human corpus and \citet{min2023beyond} argue that IdentityChain measures a distinct quality from conventional accuracy. Instead, we have shown that \RTC is strongly correlated with conventional accuracy for a given benchmark, does not require human annotated examples, and covers multiple tasks.

\textbf{Back Translation}
At a high-level, the forward and backward samples in \RTC resemble back translation -- a data augmentation technique for machine translation~\citep{sennrich2016improving, edunov2018understanding,sugiyama2019data}.
Back translation commonly does not enforce semantic equivalence and may result in noisy data, when back translation is used to perform data augmentation at scale, it can still be very useful for training since models are robust to some level of noise. Instead our focus is model evaluation.

\textbf{Code Synthesis Benchmarks}
The most common benchmarks for code synthesis include HumanEval~\cite{chen2021evaluating}, MBPP~\cite{austin2021program}, APPS~\cite{hendrycks2021measuring}, and CodeContests~\cite{li2022competition}. These benchmarks have a curated set of input (natural language) specifications and test cases for each example, and this is non-trivial to expand. In contrast to these benchmarks, \RTC can fairly easily be expanded to include new domains (\autoref{subsec:rtc domain variance}).

Finally, the HumanEvalExplain task of HumanEvalPack~\citep{muennighoff2023octopack} is a special case of \synthesisrtc without forward/backward sampling for HumanEval. In contrast to that benchmark, we acknowledge that tasks like HumanEvalExplain evaluate both synthesis and code description, make them more robust through sampling, and introduce new tasks. Furthermore, we show that \synthesisrtc tasks do actually correlate with widely accepted metrics and thus are worth measuring.

\textbf{Faithfulness}
\citet{atanasova-etal-2023-faithfulness} proposed evaluating the faithfulness of natural language explanations based on a model's ability to generate the same output if the explanation is included in the input. The consistency between the outputs are evaluated, since a faithful explanation should guide the model in generating the same output. However, the output does not necessarily have to be correct, and the explanation does not have to be descriptive. The main requirement is that the explanation does not include any information that might lead the model to generate a different output. In contrast, \RTC evaluates the correctness between the input and the prediction from the backward pass, and necessitates good predictions in both directions.

\section{Discussion \& Conclusions}
In this work we used the concept of round-trip correctness for model evaluation and found that \RTC strongly correlates with performance on narrow-domain human-curated benchmarks, measuring a similar quality of a model's performance.
\RTC allows us to complement existing narrow-domain human-annotated benchmarks and measure an LLM's performance on a much wider spectrum of domains.
\RTC allows us to expand our evaluations into new tasks such as code and edit description, and code editing without requiring human annotations.

We hope that this work motivates the community towards expanding the code evaluation benchmarks beyond narrow-domain ones.
\RTC seems to strike a balance between correlating with existing metrics and allowing to expand to new domains and tasks.
Thus we recommend to complement existing benchmarks with \RTC using a strong $\similarity(\cdot)$ to expand the breadth of the evaluated domains and tasks.
We note that \RTC should be complemented with good qualitative understanding of the LLM's error modes and suggestions rather than used blindly as a metric to be maximized.

\section*{Impact Statement}
This paper presents work whose goal is to advance the evaluation of LLMs.
Having accurate LLM evaluations can help produce more useful models and reduce the (environmental) cost of their development.
The deployment of LLMs has many potential societal consequences, none which we feel must be specifically highlighted here.
However, it should be noted that optimizing for any single (evaluation) metric without considering the broader setting --- including any ethical concerns --- in which a machine learning model will be deployed should be avoided:
No single metric, including those presented in this work, can capture the wide range of effects --- positive or negative --- that the deployment of a model may have and hence we encourage users of \RTC to always take a holistic approach when evaluating their models.

\section*{Acknowledgements}
We would like to thank Nate Kushman, Charles Sutton, and Danny Tarlow for useful discussions.
We would also like to thank the anonymous reviewers for their constructive feedback.
\bibliographystyle{icml2024}
\bibliography{references.bib}

\newpage
\appendix
\onecolumn
\section{Open-Source Projects for \autoref{subsec:rtc domain variance}}
\label{appx:oss project list}
{\tiny
\begin{longtable}[p]{lp{5cm}} \toprule
\texttt{github.com/AliRn76/panther} & Web framework\\
\texttt{github.com/JoshData/python-email-validator} & Email validation library\\
\texttt{github.com/Ousret/charset\_normalizer} & Encoding detector \\
\texttt{github.com/PantherPy/pantherdb} & Database\\
\texttt{github.com/SimonGreenhill/treemaker} & Tree formatting \\
\texttt{github.com/Textualize/rich} & Rich text formatting\\
\texttt{github.com/TheAlgorithms/Python} & Educational algorithm implementations\\
\texttt{github.com/agronholm/typeguard} & Runtime type checking\\
\texttt{github.com/aio-libs/async-lru} & Cache for Python's asyncio\\
\texttt{github.com/akoumjian/datefinder} & Extract dates from text\\
\texttt{github.com/alexmojaki/executing} & Python execution frame inspection \\
\texttt{github.com/andialbrecht/sqlparse} & SQL parser\\
\texttt{github.com/antonagestam/phantom-types} & Runtime type annotations\\
\texttt{github.com/caesar0301/treelib} & Tree data structures\\
\texttt{github.com/casbin/pycasbin} & Authorization library\\
\texttt{github.com/chaimleib/intervaltree} & Interval tree implementation\\
\texttt{github.com/cpburnz/python-pathspec} & File path pattern matching\\
\texttt{github.com/datafolklabs/cement} & Application framework\\
\texttt{github.com/dateutil/dateutil} & Date and time utilties\\
\texttt{github.com/davidhalter/jedi} & Autocompletion and refactoring library\\
\texttt{github.com/dgasmith/opt\_einsum} & Optimizing einsum functions\\
\texttt{github.com/eigenein/protobuf} & Python ProtoBuf implementation\\
\texttt{github.com/ekzhu/datasketch} & Probabilistic Data Structures\\
\texttt{github.com/fabiocaccamo/python-benedict} & Dictionary implemntation\\
\texttt{github.com/facelessuser/soupsieve} & CSS Selector \\
\texttt{github.com/facelessuser/wcmatch} & Wilcard File Name matching library \\
\texttt{github.com/foutaise/texttable} & ASCII table creation \\
\texttt{github.com/graphql-python/graphql-core} & GraphQL port \\
\texttt{github.com/hgrecco/pint} & Measurement Unit Library\\
\texttt{github.com/hukkin/tomli} & TOML parser\\
\texttt{github.com/jab/bidict} & Bidirecitonal map data structure\\
\texttt{github.com/jaraco/jaraco.collections} & Collection data structures\\
\texttt{github.com/jaraco/path} & File system path manupulation\\
\texttt{github.com/jd/tenacity} & Retrying library\\
\texttt{github.com/jsh9/pydoclint} & Docsting linter\\
\texttt{github.com/kjd/idna} & Internationalized Domain Names library \\
\texttt{github.com/lemon24/reader} &  Feed reader library\\
\texttt{github.com/lidatong/dataclasses-json} & Dataclass serialization library\\
\texttt{github.com/magmax/python-readchar} & Library to read characters and key strokes\\
\texttt{github.com/mahmoud/boltons} & Generic Python utilities\\
\texttt{github.com/marcusbuffett/command-line-chess} & Chess in the CLI \\
\texttt{github.com/martinblech/xmltodict} & XML library\\
\texttt{github.com/mbr/asciitree} & Print trees in ASCII\\
\texttt{github.com/mckinsey/vizro} & Data visualization\\
\texttt{github.com/mgedmin/objgraph} & Inspect object graphs\\
\texttt{github.com/microsoft/lsprotocol} & Code generator for LSP \\
\texttt{github.com/mkdocstrings/griffe} & API Documentation\\
\texttt{github.com/mkorpela/overrides} & Override decorator\\
\texttt{github.com/more-itertools/more-itertools} & Iterator utilities\\
\texttt{github.com/mozilla/bleach} & HTML sanitization\\
\texttt{github.com/msiemens/tinydb} & Document database \\
\texttt{github.com/openlawlibrary/python-docx} & DocX library \\
\texttt{github.com/pallets/click} & CLI interface \\
\texttt{github.com/pallets/flask} & Web app framework\\
\texttt{github.com/pavdmyt/yaspin} & Terminal spinner \\
\texttt{github.com/pydantic/pydantic} & Data validation library\\
\texttt{github.com/pydata/patsy} & Statistical Model Description\\
\texttt{github.com/pydata/xarray} & n-D arrays\\
\texttt{github.com/pygments/pygments} & Code highlighting\\
\texttt{github.com/pyparsing/pyparsing} & PEG parsers\\
\texttt{github.com/python-jsonschema/jsonschema} & JSON Schema library \\
\texttt{github.com/pytoolz/toolz} & Functional utilities \\
\texttt{github.com/pyupio/safety} & Vulnerability Detection \\
\texttt{github.com/serge-sans-paille/beniget} & Static Python analysis\\
\texttt{github.com/simonw/datasette} & Tool for data exploration and publishing \\
\texttt{github.com/sqids/sqids-python} & Short Unique Ids library\\
\texttt{github.com/sybrenstuvel/python-rsa} & RSA implementation\\
\texttt{github.com/tartley/colorama} & Colored terminal text\\
\texttt{github.com/tiangolo/fastapi} & Web framework\\
\texttt{github.com/tqdm/tqdm} & Progress bar library \\
\texttt{gitlab.com/ericvsmith/toposort} & Topological sorting library\\
\texttt{launchpad.net/code/beautifulsoup} & HTML and XML parsing \\\bottomrule
\end{longtable}
} 

\section{Lift on Diverse Projects}
\label{appx:lift plot}
\begin{figure*}
\centering
\includegraphics[trim={0cm 0cm 0cm 0cm},clip,width=\textwidth]{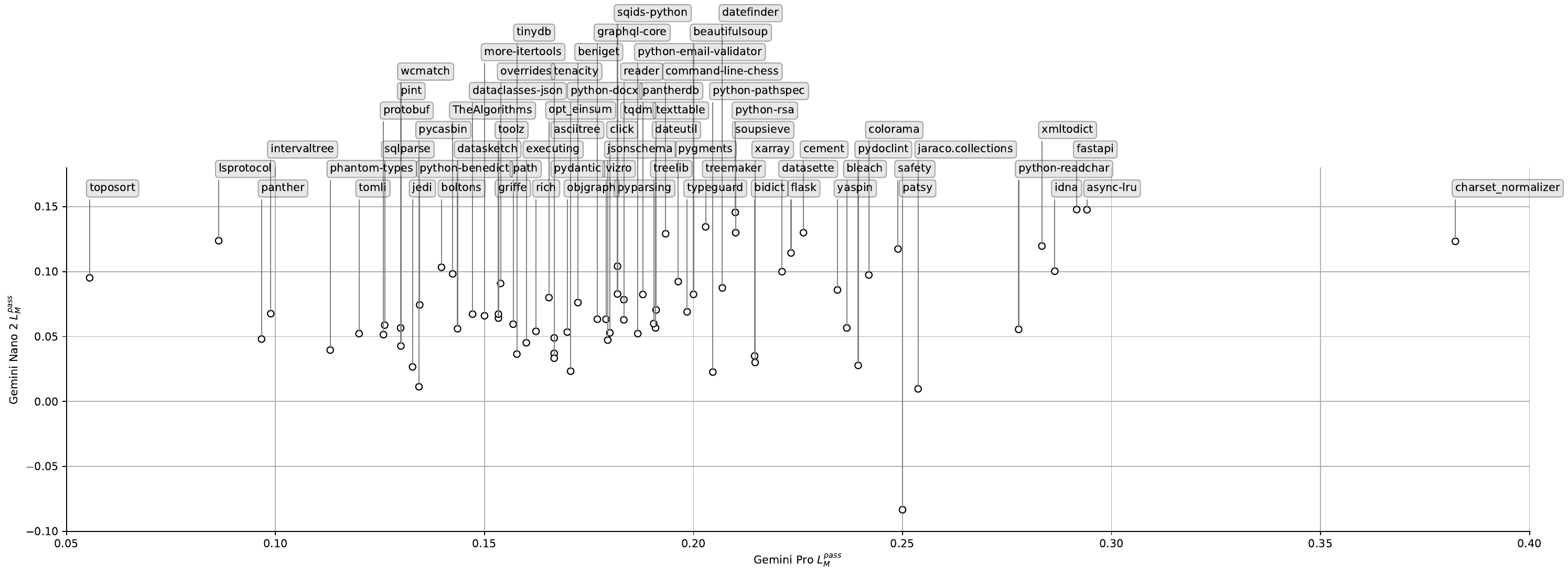}
    \caption{Lift}
    \label{fig:lift circus}
\end{figure*}

\autoref{fig:lift circus} show a plot of the lift \lift for each project.
Again, we see  a wide variability of the lift across different projects.

\section{\editingrtc Examples}
\label{app:editingrtc_examples}
In this section, we provide several qualitative \editingrtc examples. In Example \#1, all three edit descriptions sampled in the forward pass lead to edits which exactly match the original in the backward pass, consequently achieving a perfect \RTC score of 100.0. This is possible because the underlying model is able to generate high-quality descriptions that capture the essence of the edit in the forward pass and also precisely perform the described edit in the backward pass. This demonstrates how we can reliably evaluate a model without a labeled evaluation set. 

Nonetheless, in our setting, we can consider the PR comments in the CodeReviewer test set as ground truth edit descriptions. For this particular example, we find that leveraging the PR comment in the backward pass (in place of the sampled descriptions) does lead to the model predicting the matching edit. With a supervised evaluation setup, we can assess the quality of the sampled edit descriptions by comparing to the ground truth edit description with a standard text generation metric like BLEU. Here, we find that irrespective of whether the BLEU score is high (\eg Sample \#3) or low (\eg Sample \#2), the descriptions sufficiently describe the edit in such a way that the model can re-derive the edit using these descriptions in the backward pass. While the average description BLEU is still relatively high in Example \#1, we find that it is often much lower even when the predicted descriptions accurately reflect the edit, such as in Example \#2. 

The previous examples pertain to instances in which the \RTC score is high. We now consider cases in which the \RTC score is low. In Example \#3, none of the sampled descriptions accurately reflect the edit, and consequently, they do not lead to the matching edit in the backward pass. While the underlying model is able to generate an edit which is consistent with the description for Sample \#3, the descriptions for Sample \#1 and Sample \#2 are too confusing for the model such that it simply copies the old code without performing any edits. Despite the fact that these descriptions are inaccurate, the BLEU score (with respect to the PR comment) is relatively high, especially for Sample \#1, again highlighting another instance in which \RTC serves as a more reliable evaluation metric.

In Example \#4, we have an example in which the \RTC score is 33.333. Of the three predicted descriptions, only one accurately describes the edit (Sample \#1), which is also the only one which leads to a matching edit in the backward pass. The predicted edit for the description in Sample \#2 is partially consistent with the incorrect description as it does remove \textit{schedule} from \textit{wp\_cron\_schedule\_https\_detection}. For Sample \#3, the model is likely confused by the reference to \textit{wp\_https\_detection} in the predicted description, since it does not exist in the code, and consequently it simply copies the old code without performing any edits. Interestingly, the predicted descriptions in Sample \#1 and Sample \#3 achieve similar BLEU scores with respect to the PR comment, despite the fact that the first one is accurate while the third one is not. On the other hand, the first description leads to a matching edit during the backward pass while the third one does not.

\begin{longtable}{lcc}
\toprule
\textbf{Example \#1} && \\
\hline
\begin{lstlisting}
Describe concisely and accurately with natural language the differences
between the old and new code shown below.
[old]
 }
 cmd := []string{"go", "build"}
 if ctx.Config.Build.Flags != "" {
    cmd = append(cmd, (*@\hlc[pink]{strings.Split(ctx.Config.Build.Flags, " ")}@*)...)}
}
cmd = append(cmd, "-ldflags="+ldflags, "-o", output, ctx.Config.Build.Main)
if err := run(goos, goarch, cmd); err != nil {
 
[new]
 }
cmd := []string{"go", "build"}
if ctx.Config.Build.Flags != "" {
    cmd = append(cmd, (*@\hlc[green]{strings.Fields(ctx.Config.Build.Flags)}@*)...)
 }
cmd = append(cmd, "-ldflags="+ldflags, "-o", output, ctx.Config.Build.Main)
if err := run(goos, goarch, cmd); err != nil {

[edit description]

\end{lstlisting}& \\
\midrule
\bf Description & \bf Desc BLEU ($\rightarrow$) & \bf Edit Exact Match ($\leftarrow$) \\
\hline
\small{\textbf{PR Comment:} Is`strings.Fields(ctx.Config.Build.Flags)` better?} & - & \cmark \\
\small{\textbf{Baseline:} Edit.}  &  - &  \xmark \\
\hline
\small{\textbf{\editingrtc Forward Samples:}} && \\
\makecell[l]{\small{\textbf{\#1:}  Please switch to strings.Fields() and keep the code}\\\small{consistent with above.} }
& 15.917 & \cmark \\
\small{\textbf{\#2:} Please use correct strings.Fields call.} & 7.171 & \cmark \\
\makecell[l]{\small{\textbf{\#3:} This should use a Go-style append of}\\\small{`strings.Fields(ctx.Config.Build.Flags)` to cmd.}}
 & 50.868 & \cmark\\
\midrule
& \bf Avg Desc BLEU & \bf \rtcexmatch  \\
\hline
& 24.652 & 100.0 \\
\bottomrule
&& \\

\toprule
\textbf{Example \#2} && \\
\hline
\begin{lstlisting}
Describe concisely and accurately with natural language the differences
between the old and new code shown below.

[old]
 }
// check if files were uploaded through the manifest
if (*@\hlc[pink]{dirManifest.Size())}@*) == 0 {
 return swarm.ZeroAddress, fmt.Errorf("no files added from tar")
}

[new]
 }
// check if files were uploaded through the manifest
if (*@\hlc[green]{dirManifest.Length())}@*) == 0 {
 return swarm.ZeroAddress, fmt.Errorf("no files added from tar")
}

[edit description]

\end{lstlisting}& \\
\midrule
\bf Description & \bf Desc BLEU ($\rightarrow$) & \bf Edit Exact Match ($\leftarrow$) \\
\hline
\makecell[l]{\small{\textbf{PR Comment:} I think `Length` is more legible here as`Size` might}\\\small{infer actual size in bytes}}
 & - & \cmark \\
\small{\textbf{Baseline:} Edit.}  &  - &  \xmark \\
\hline
\small{\textbf{\editingrtc Forward Samples:}} && \\
\makecell[l]{\small{\textbf{\#1:} please use the length ( Length() ) method instead of}\\\small{the size() method.}}
 & 2.416 & \cmark \\
\makecell[l]{\small{\textbf{\#2:} In this case, "Size" is not a good thing to read.}\\\small{"Length" is better.}}
 & 2.390 & \cmark \\
\makecell[l]{\small{\textbf{\#3:} Please use Length() instead of Size() as this code}\\\small{will work more correctly across different go versions.}} & 2.443 & \cmark\\
\midrule
& \bf Avg Desc BLEU & \bf \rtcexmatch   \\
\hline
& 2.416 & 100.0 \\
\bottomrule
&& \\

\toprule
\textbf{Example \#3} && \\
\hline
\begin{lstlisting}
Describe concisely and accurately with natural language the differences
between the old and new code shown below.

[old]
kms_master_key_id = long_uid()
sse_specification = {"Enabled": True, "SSEType": "KMS",
    "KMSMasterKeyId": kms_master_key_id}
(*@\hlc[pink]{kms\_master\_key\_arn = "arn:aws:kms:\%s:\%s:key/\%s" \% (}@*)
    (*@\hlc[pink]{aws\_stack.get\_local\_region(),}@*)
    (*@\hlc[pink]{TEST\_AWS\_ACCOUNT\_ID,}@*)
    (*@\hlc[pink]{kms\_master\_key\_id,}@*)
(*@\hlc[pink]{)}@*)
result = dynamodb.create_table(
    TableName=table_name,
    
[new]
kms_master_key_id = long_uid()
sse_specification = {"Enabled": True, "SSEType": "KMS",
    "KMSMasterKeyId": kms_master_key_id}
(*@\hlc[green]{kms\_master\_key\_arn = aws\_stack.kms\_key\_arn(kms\_master\_key\_id)}@*)
result = dynamodb.create_table(
    TableName=table_name,

[edit description]

\end{lstlisting}& \\
\midrule
\bf Description \& Edit & \bf Desc BLEU ($\rightarrow$) & \bf Edit Exact Match ($\leftarrow$) \\
\hline
\makecell[l]{\small{\textbf{PR Comment:} Same as above - we can use}\\\small{`aws\_stack.kms\_key\_arn(..)` here.}}
 & - & \\
\begin{lstlisting}
[new]
kms_master_key_id = long_uid()
sse_specification = {"Enabled": True, "SSEType": "KMS",
    "KMSMasterKeyId": kms_master_key_id}
(*@\hlc[green]{kms\_master\_key\_arn = aws\_stack.kms\_key\_arn(kms\_master\_key\_id)}@*)
result = dynamodb.create_table(
    TableName=table_name,
\end{lstlisting}& & \cmark \\
\hline
\small{\textbf{Baseline:} Edit.}  &  - & \\
\begin{lstlisting}
[new]
kms_master_key_id = long_uid()
sse_specification = {"Enabled": True, "SSEType": "KMS",
    "KMSMasterKeyId": kms_master_key_id}
kms_master_key_arn = "arn:aws:kms:%s:%s:key/%s" % (
    aws_stack.get_local_region(),
    TEST_AWS_ACCOUNT_ID,
    kms_master_key_id,
)
result = dynamodb.create_table(
    TableName=table_name,
\end{lstlisting}&  &  \xmark \\
\hline
\makecell[l]{\small{\textbf{\editingrtc Sample \#1:} Please rename the aws\_stack function.}\\\small{'kms\_key\_arn' is the convention for methods}}
 & 20.829 & \\
\begin{lstlisting}
[new]
kms_master_key_id = long_uid()
sse_specification = {"Enabled": True, "SSEType": "KMS",
    "KMSMasterKeyId": kms_master_key_id}
kms_master_key_arn = "arn:aws:kms:%s:%s:key/%s" % (
    aws_stack.get_local_region(),
    TEST_AWS_ACCOUNT_ID,
    kms_master_key_id,
)
result = dynamodb.create_table(
    TableName=table_name,
\end{lstlisting} & & \xmark \\
\hline
\makecell[l]{\small{\textbf{\editingrtc Sample \#2:} Avoid calling into other modules for simple}\\\small{formatting operations. The kms\_key\_arn operation in aws\_stack is a one-liner,}\\\small{so no need to extract it as its own function. Similarly, avoid calling}\\\small{AWS APIs on the module level if you need to pass parameters.}}
 & 9.344 &  \\
\begin{lstlisting}
[new]
kms_master_key_id = long_uid()
sse_specification = {"Enabled": True, "SSEType": "KMS",
    "KMSMasterKeyId": kms_master_key_id}
kms_master_key_arn = "arn:aws:kms:%s:%s:key/%s" % (
    aws_stack.get_local_region(),
    TEST_AWS_ACCOUNT_ID,
    kms_master_key_id,
)
result = dynamodb.create_table(
    TableName=table_name,
\end{lstlisting}&  & \xmark \\
\hline
\makecell[l]{\small{\textbf{\editingrtc Sample \#3:} Remove "get\_local\_region()" call from here.}\\\small{This should be using the region from the AWS stack.}}
 &  5.066 & \\
\begin{lstlisting}
[new]
kms_master_key_id = long_uid()
sse_specification = {"Enabled": True, "SSEType": "KMS",
    "KMSMasterKeyId": kms_master_key_id}
kms_master_key_arn = "arn:aws:kms:%s:%s:key/%s" % (
    (*@\hlc[green]{aws\_stack.get\_region(),}@*)
    TEST_AWS_ACCOUNT_ID,
    kms_master_key_id,
)
result = dynamodb.create_table(
    TableName=table_name,

\end{lstlisting}&  & \xmark \\
\midrule
& \bf Avg Desc BLEU & \bf \rtcexmatch   \\
\hline
&  11.746 &  0.000 \\
\bottomrule
&& \\

\toprule
\textbf{Example \#4} && \\
\hline
\begin{lstlisting}
Describe concisely and accurately with natural language the differences
between the old and new code shown below.

[old]
 }
// HTTPS detection.
add_action( 'init', (*@\hlc[pink]{'wp\_cron\_schedule\_https\_detection'}@*) );
add_action( 'wp_https_detection', 'wp_update_https_detection_errors' );
add_filter( 'cron_request', 'wp_cron_conditionally_prevent_sslverify', 9999 );

[new]
 }
// HTTPS detection.
add_action( 'init', (*@\hlc[green]{'wp\_schedule\_https\_detection'}@*) );
add_action( 'wp_https_detection', 'wp_update_https_detection_errors' );
add_filter( 'cron_request', 'wp_cron_conditionally_prevent_sslverify', 9999 );

[edit description]

\end{lstlisting}& \\
\midrule
\bf Description \& Edit & \bf Desc BLEU ($\rightarrow$) & \bf Edit Exact Match ($\leftarrow$) \\
\hline
\makecell[l]{\small{\textbf{PR Comment:} Would this be better as `admin\_init` instead?}\\\small{This will reduce frontend DB writes.}\\\small{ ```suggestion add\_action( 'admin\_init',}\\\small{ 'wp\_cron\_schedule\_https\_detection' ); ```}}
 & - & \\
\begin{lstlisting}
[new]
// HTTPS detection.
add_action( (*@\hlc[green]{'admin\_init'}@*), 'wp_cron_schedule_https_detection');
add_action( 'wp_https_detection', 'wp_update_https_detection_errors' );
add_filter( 'cron_request', 'wp_cron_conditionally_prevent_sslverify', 9999 );
\end{lstlisting}& & \xmark \\
\hline
\small{\textbf{Baseline:} Edit.}  &  - & \\
\begin{lstlisting}
[new]
 }
// HTTPS detection.
add_action( 'init', 'wp_cron\_schedule_https_detection' );
add_action( 'wp_https_detection', 'wp_update_https_detection_errors' );
add_filter( 'cron_request', 'wp_cron_conditionally_prevent_sslverify', 9999 );
\end{lstlisting}&  &  \xmark \\
\hline
\makecell[l]{\small{\textbf{\editingrtc Sample \#1:} The filter/hook name used to schedule events}\\\small{changed from `wp\_cron\_schedule\_` to just `wp\_schedule\_`.}\\\small{Remove the `wp\_cron\_prefix` here.}}
 & 11.667 & \\
\begin{lstlisting}
[new]
 }
// HTTPS detection.
add_action( 'init', (*@\hlc[green]{'wp\_schedule\_https\_detection'}@*) );
add_action( 'wp_https_detection', 'wp_update_https_detection_errors' );
add_filter( 'cron_request', 'wp_cron_conditionally_prevent_sslverify', 9999 );
\end{lstlisting} & & \cmark \\
\hline
\makecell[l]{\small{\textbf{\editingrtc Sample \#2:} Please replace "schedule" with "cron"; this}\\\small{causes confusion to those who are unfamiliar with Wordpress jargon.}}
 & 0.757 &  \\
\begin{lstlisting}
[new]
 }
// HTTPS detection.
add_action( 'init', (*@\hlc[green]{'wp\_cron\_https\_detection'}@*) );
add_action( 'wp_https_detection', 'wp_update_https_detection_errors' );
add_filter( 'cron_request', 'wp_cron_conditionally_prevent_sslverify', 9999 );
\end{lstlisting}&  & \xmark \\
\hline
\makecell[l]{\small{\textbf{\editingrtc Sample \#3:} Please rename the actions and filter used in this}\\\small{commit from `wp\_https\_detection` to `wp\_cron\_conditionally\_prevent\_sslverify`}\\\small{to `wp\_https\_detection`.}}
 &  11.345 & \\
\begin{lstlisting}
[new]
 }
// HTTPS detection.
add_action( 'init', 'wp_cron\_schedule_https_detection' );
add_action( 'wp_https_detection', 'wp_update_https_detection_errors' );
add_filter( 'cron_request', 'wp_cron_conditionally_prevent_sslverify', 9999 );
\end{lstlisting}& & \xmark \\
\midrule
& \bf Avg Desc BLEU & \bf \rtcexmatch   \\
\hline
&  7.923 & 33.333  \\
\bottomrule
\end{longtable}

\section{Length Statistics}
\label{appx:length stats}
\begin{figure}
\centering
\includegraphics[scale=0.6]{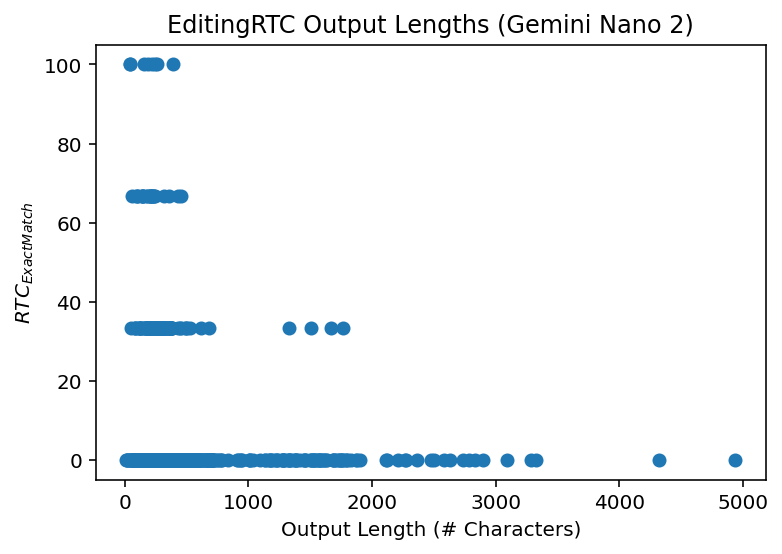}
    \caption{Relationship between output length and \rtcexmatch for \editingrtc.}
    \label{fig:editing_rtc_output_lengths}
\end{figure}

For \editingrtc, we find that the outputs which achieve a non-zero \rtcexmatch score are generally shorter than the outputs that achieve an \rtcexmatch score of zero (\autoref{fig:editing_rtc_output_lengths}). For instance, with Gemini Nano 2, the average number of characters for outputs achieving non-zero scores is 302.6 while it is 454.2 for outputs which have \rtcexmatch scores of zero. This is intuitive since it becomes more difficult to exactly match the input as the length increases.

\end{document}